\documentclass[aps,prl,showpacs,floatfix,twocolumn,letterpaper]{revtex4}
\usepackage[pdftex]{graphicx}
\usepackage{amsmath}
\usepackage{hyperref}
\usepackage{latexsym}
\usepackage{color}
\usepackage{amssymb}
\usepackage{amsmath}
\usepackage{float}
\usepackage[normalem]{ulem}   
\usepackage{times}

\usepackage{graphicx}

\usepackage{soul,xcolor}


\begin{document}
\setstcolor{blue}

\title{Milne's Equation revisited: Exact Asymptotic Solutions}

\author{D. Shu, I. Simbotin, R. C\^ot\'e}
\affiliation{Department of Physics, University of Connecticut, Storrs, CT 06268, USA}


\begin{abstract}
We present novel approaches for solving Milne's equation, which was introduced
in 1930 as an efficient numerical scheme for the Schr\"odinger
equation. Milne's equation appears in a wide class of physical
problems, ranging from astrophysics and cosmology, to quantum
mechanics and quantum optics. We show how a third order linear
differential equation is equivalent to Milne's non-linear equation,
and can be used to accurately calculate Milne's amplitude and phase
functions. We also introduce optimization schemes to achieve a
convenient, fast, and accurate computation of wave functions using an
economical parametrization. These new optimization procedures answer
the long standing question of finding non-oscillatory solutions of
Milne's equation. We apply them to long-range
potentials and find numerically exact asymptotic solutions.
\end{abstract}

\maketitle


In 1930, Milne introduced a phase-amplitude approach \cite{Milne_origin}
for tackling time-independent wave equations.  The resulting equation
for the amplitude $y(x)$, known as Milne's equation, is a nonlinear
differential equation of second order,
\begin{equation}\label{eq:Milne-original}
\partial_{x}^2y=U (x ) y  +\frac{q^2}{y^3 },
\end{equation}
where $U(x)$ is a function of the independent variable $x$, and $q$ is
a constant.  Replacing a linear wave equation with a nonlinear one for
the amplitude is a price worth paying, as Milne's approach can yield a
very economical parametrization of highly oscillating wave functions.
In addition, Eq.~(\ref{eq:Milne-original}) allows for novel
formalistic approaches, and is now used in many areas in physics, as
well as being investigated for its mathematical properties
\cite{discrete_Pinney}.

In astrophysics, studies of stellar equilibrium for white dwarfs or
neutron stars employ certain transformations to rewrite the
Tolman-Oppenheimer-Volkoff equations in the form of Milne's equation,
which is more amenable for analyzing the stability of stellar objects
\cite{TOV_neutron_star}. Milne's equation is also used in scalar field
cosmologies for deriving exact cosmological models \cite{ekpyrotic},
for establishing a dynamical correspondence between certain types of
cosmologies and quasi-two-dimensional BEC \cite{BEC_cosmology}, or to
investigate the quantum effects of relic gravitons
\cite{Relic_Gravitons}. Also, the problem of the quantized motion for
free fall in a gravitational potential \cite{quantum_freefall} has
been investigated using Milne's equation.  Moreover, Milne's approach
is used in conjunction with a simple gauge model for studying bilayer
graphene \cite{Bilayer_graphene}. Recent work on Berry's phase in
arbitrary dimensions \cite{berry_phase_d_dimension}, as well as
studies of the propagation of quantized electromagnetic waves through
nondispersive media \cite{Pedrosa_QEM_wave,Pedrosa_PRA} also make use
of Milne's equation.

In this Letter, we consider the original problem that led Milne to
develop his equation, namely the one-dimensional differential
equation,
\begin{equation}
 \psi''(x) - U(x) \psi(x)=0.
 \label{eq:DE-original}
\end{equation}
By using two linearly independent solutions $u(x)$ and $v(x)$ of
Eq.~(\ref{eq:DE-original}), and defining the function $y(x)=
\sqrt{u^2+v^2}$, Milne showed \cite{Milne_origin} that
Eq.~(\ref{eq:DE-original}) leads to the differential equation
(\ref{eq:Milne-original}) where the constant $q = u'v-uv'$ is the
Wronskian of $u$ and $v$.  We emphasize that, assuming $q\ne0$, any
particular solution $y(x)$ of Eq.~(\ref{eq:Milne-original}) can be
used to construct the most general solution of
Eq.~(\ref{eq:DE-original}) via the simple parametrization,
\begin{equation}
  \psi(x) = C y(x) \sin \left[ \theta(x) -\theta_0\right], \quad
  \theta(x) = q\!\int_{x_0}^x \frac{d\tau}{y^2(\tau)},\\
  \label{eq:Milne-original-solution-y}
\end{equation}
where $C$ and $\theta_0$ are arbitrary constants.

Milne's phase-amplitude method has been used extensively in atomic and
molecular physics problems and also in physical chemistry
\cite{quantum_chem_Light,Multiple_well_Soo_Yin}.  It is especially
suitable in the multi-channel quantum defect theory framework
\cite{Raoult_AND_Mies,MQDT_Feshbach_osseni,Greene_fig,Greene_John_PRL81,
  phase_amp_John_PRA49,MQDT_Julienne,John_and_Paul} because it makes
it possible to construct optimal reference functions.  We note that
this simple parametrization of the wave function has been employed in
other approaches, {\it e.g.}, the WKB approximation and the variable
phase method \cite{calogero}. The latter is an exact approach, as is
Milne's. However, Milne's approach is more advantageous, e.g., in the
context of quantum defect theory, see Greene et
al.~\cite{Chris_Greene_QDT_milne_formalism}.

Although Milne's method is very appealing, it presents some
challenges.  First, Eq.~(\ref{eq:Milne-original}) is obviously
nonlinear, though tractable numerically. Second, a more serious
difficulty is finding the elusive optimally smooth solution among the
infinite number of oscillatory solutions of
Eq.~(\ref{eq:Milne-original}), which was the original goal of Milne,
i.e., replacing a rapidly oscillating wave function $\psi(x)$ by two
slowly varying functions $y(x)$ and $\theta(x)$.  As mentioned above,
any solution $y(x)$ and its associated phase $\theta(x)$ can be used
in Eq.~(\ref{eq:Milne-original-solution-y}) to parametrize $\psi(x)$,
but an oscillatory behavior of $y(x)$ would render this approach
inconvenient and computationally expensive.  Numerous attempts have
been made to find the optimal solution for Milne's equation, yet they
are either insufficiently optimal due to reliance on the WKB
approximation
\cite{Chris_Greene_QDT_milne_formalism,second_order_WKB_wf_correction,sidky_phys_essay},
not robust enough with respect to iterations of WKB
\cite{Seaton_and_peach,George}, or not general enough
\cite{limited_maximum_optimization}.  In the remainder of the Letter,
we present two novel optimization schemes which are efficient,
convenient, accurate, and general.

\textit{ Linear differential equation for the envelope ---} To remove
the non-linear term in Milne's equation, we introduce a new quantity,
namely $\rho(x) \equiv y^2(x)$, which we refer to as the envelope
function. Simply substituting $y=\sqrt{\rho}$ in
Eq.~(\ref{eq:Milne-original}), we obtain
\begin{equation}
 \rho '' - 2U\rho -\frac{2}{\rho} 
 \left[ q^2 + \frac{1}{4} \left( \rho '\right)^2 \right] =0 \;.
 \label{eq:Milne-original-2nd-linear}
\end{equation}
Next, we multiply Eq.~(\ref{eq:Milne-original-2nd-linear}) with $\rho$
and take the derivative
\cite{Japanese_linear,discrete_Pinney,nist_bessel_hand_book}, yielding
\begin{equation}
    \rho ''' - 4U\rho' - 2U'\rho =0 \;,
    \label{eq:Milne-original-3rd-linear}
\end{equation}
which is a linear differential equation, albeit of third order.  We
emphasize that, despite the obvious advantages of replacing Milne's
nonlinear equation (\ref{eq:Milne-original}) with the linear equation
(\ref{eq:Milne-original-3rd-linear}), difficulties in finding the
non-oscillatory envelope persist.  Thus, an optimization procedure is
needed to isolate the smooth solution, which will provide the initial
conditions of Eq.~(\ref{eq:Milne-original-3rd-linear}) for propagating
$\rho(x)$ along the $x$-axis, with a numerical solver employing a
spectral integration method based on Chebyshev polynomials
\cite{El-gendi,Greengard,Mihaila}.

\begin{figure}[b]
\includegraphics[width=1.0\columnwidth]{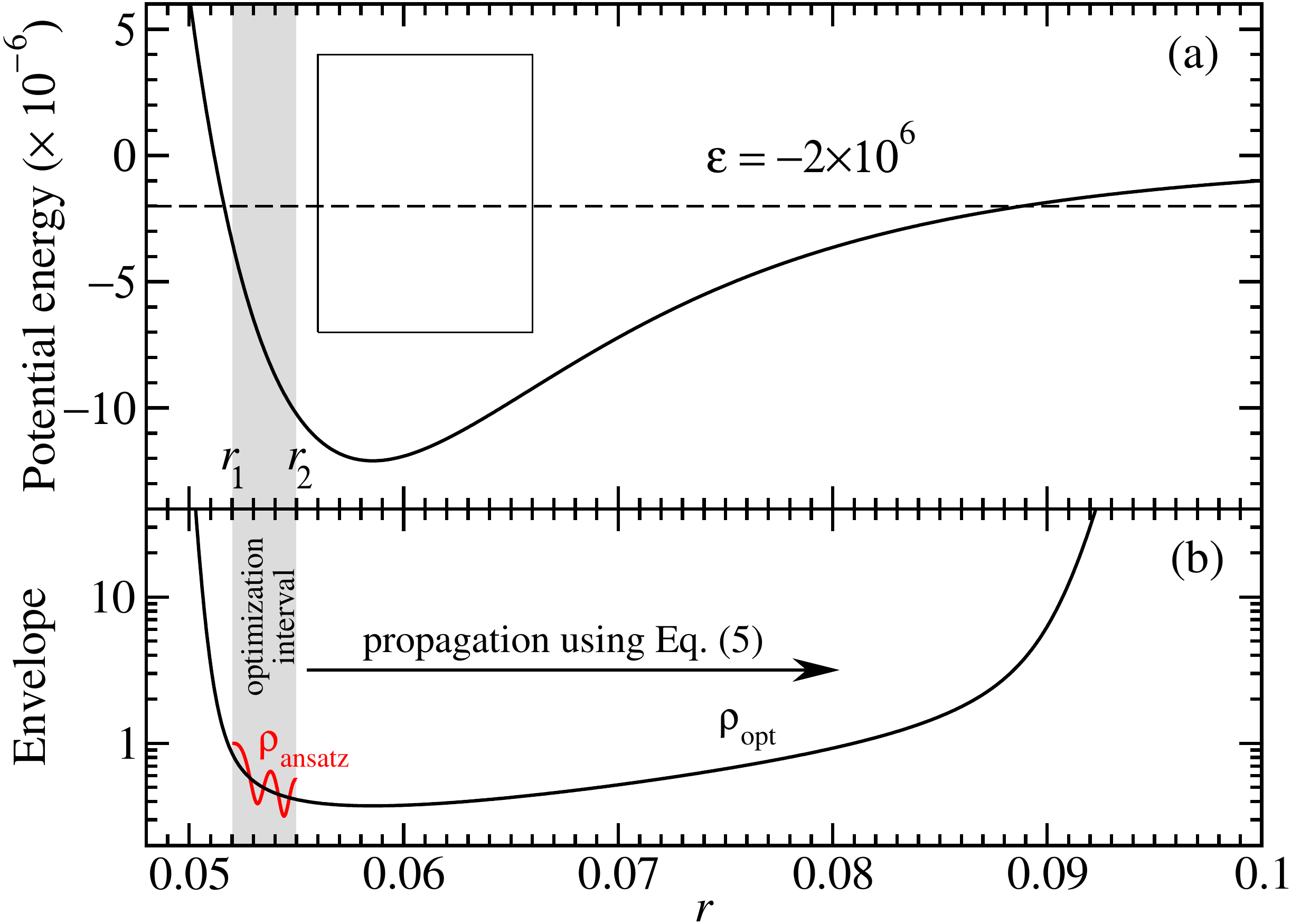}
\caption{\label{fig:scheme}  (a) Potential energy, Eq.~(\ref{eq:VCs2}), used for optimization. Rectangle corresponds
the ranges of $r$ and $\varepsilon$ in Fig.~\ref{fig:surf}.
Shaded area marks the optimization interval.     (b) Red line for
$\rho_{\mathrm{ansatz}}=\phi^2+\chi^2$, corresponding to $A=B=1,\, C=0$, and black line for 
$\rho_{\mathrm{opt}}=A\phi^2+B\chi^2+2C\phi\chi$ with $A=0.8533850906254,\, B=1.245534003812,\, C=-0.2508388899674$, which was propagated outside the optimization interval using Eq.~(\ref{eq:Milne-original-3rd-linear}).  }
\end{figure}

\textit{ Optimization ---} We now introduce a simple optimization
method for finding the smooth envelope in any classically allowed
region.  First we note that the solutions $u$ and $v$ in the original
expression $y = \sqrt{u^2+v^2}$ can each be written as superpositions
of other independent solutions $\phi(x)$ and $\chi(x)$, i.e.,
$u=a\phi+b\chi$ and $v=\alpha\phi+\beta\chi$.  Thus we obtain
\begin{equation} 
   \rho  =  A\phi^2 + B\chi^2 + 2C\phi\chi \; , \label{eq:rho-ABC}
\end{equation}
with $A=a^2+\alpha^2$, $B=b^2+\beta^2$, and
$C=ab+\alpha\beta$. Similarly, the Wronskian $q=u'v-uv'$ yields
\begin{equation}
q^2  =  (\phi\chi'-\chi\phi')^2(AB-C^2)\;. \label{eq:q2-ABC}
\end{equation}
If $\phi$ and $\chi$ are two linearly independent solutions of
Eq.~(\ref{eq:DE-original}), it can be shown that $\phi^2$, $\chi^2$,
and $\phi\chi$ are three linearly independent solutions of
Eq.~(\ref{eq:Milne-original-3rd-linear}).  Thus, if $A,\,B,\,C$ are
regarded as arbitrary constants, Eq.~(\ref{eq:rho-ABC}) represents the
most general solution of Eq.~(\ref{eq:Milne-original-3rd-linear}).

\begin{figure}[b]
\includegraphics[width=1.0\columnwidth]{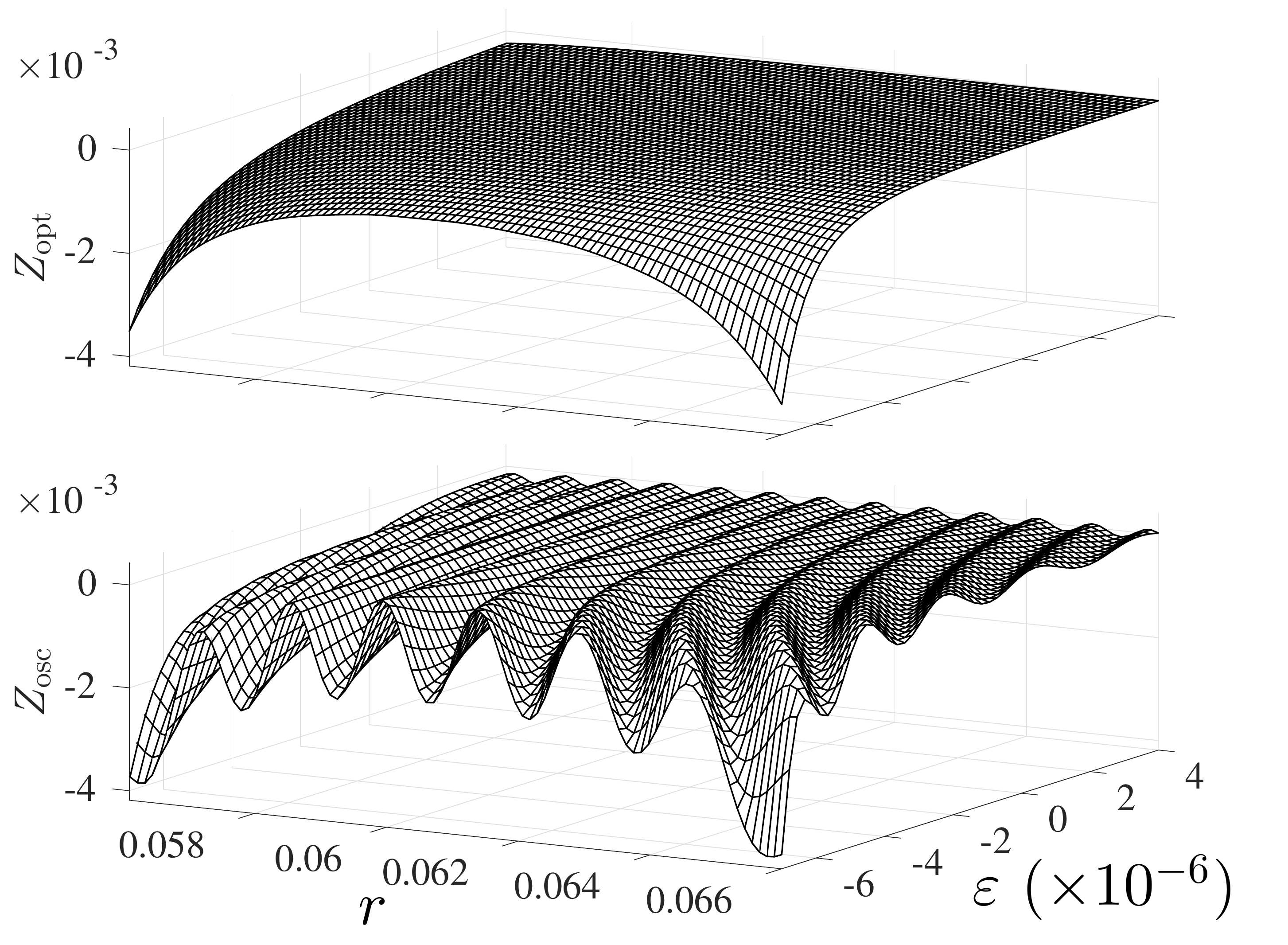}
\caption{$Z(\varepsilon,r)$ using optimization (upper) and using the
  WKB-initialization (lower).  See text for details. The range of the
  surface plot corresponds to the rectangle in
  Fig.~\ref{fig:scheme}a. }\label{fig:surf}
\end{figure}

We first compute two linearly independent solutions $\phi$ and $\chi$
of Eq.~(\ref{eq:DE-original}) within a narrow interval $[x_1,x_2]$
inside a classically allowed region.  Note that, for the optimization
to be effective, the interval $[x_1,x_2]$ should cover one or two
oscillations of $\phi$ and $\chi$.  Moreover, by computing $\phi(x)$
and $\chi(x)$ with high accuracy, we ensure that $\rho(x)$ in
Eq.~(\ref{eq:rho-ABC}) is numerically exact within $[x_1,x_2]$, and we
label it $\rho_{\rm exact}(x)$.  We emphasize that for arbitrary
parameters $A,\,B,\,C$, the envelope will be oscillatory. Next, we
expand the exact envelope over the interval $[x_1,x_2]$ in a
\emph{small} polynomial basis of size $N\lessapprox 20$, such that it
cannot reproduce the oscillations of $\rho_{\rm exact}$, in effect
downgrading it to an approximate envelope,
$\rho_{\mathrm{approx}}$.  For arbitrary parameters $(A,B,C)$, the
latter will be a poor approximation, i.e., the error
$\delta\rho=\lVert\rho_\mathrm{exact}-\rho_{\mathrm{approx}}\rVert$
will be large.  However, when the parameters are precisely optimized
such that the envelope has a non-oscillatory behavior, the error
$\delta\rho$ vanishes because the small basis is sufficient to
accurately reproduce the smooth envelope. A standard optimization
subroutine is used to minimize the error function,
\begin{equation}
 \delta \rho \equiv \underset{x\in [x_1,x_2]}{\rm max}\big| \rho_{\rm
   exact}(Q;x) -\rho_{\mathrm{approx}}(Q;x) \big|,
 \label{eq:delta-rho}
\end{equation}
over the parameter space $Q\equiv (A,\,B,\,C|q)$. We remark that
Eq.~(\ref{eq:q2-ABC}) represents a constraint for $A,\,B,\,C$, because
$q$ is assumed fixed, and thus the parameter space $Q\equiv
(A,\,B,\,C|q)$ is only two dimensional.  Our implementation uses
Chebyshev polynomials, $T_n(x)=\cos[n\, \mathrm{acos}(x)]$, for the
interpolation $\rho_{\mathrm{approx}} = \sum_{n=0}^{N-1} c_n(Q)
T_n(x)$.  The optimized (smooth) envelope is then used as an initial
condition for Eq.~(\ref{eq:Milne-original-3rd-linear}), and $\rho(x)$
is propagated on both sides of the initial working interval, to cover
the entire $x$-domain.

As a first application, we consider the time-independent radial
Schr\"odinger equation (\ref{eq:DE-original}), with $0<x<\infty$ and
$U(x)= \frac{2\mu}{\hbar^2} [V(x)-E] + \ell(\ell+1)/x^2$, where $\mu$
is the reduced mass, $V(x)$ the interaction potential, and
$\ell(\ell+1)/x^2$ the centrifugal term for a given partial wave
$\ell$. To simplify notations, we use dimensionless quantities,
$r=x/x_{s}$ and $\varepsilon = E/E_{s}$ with $E_{s}=\hbar/2\mu
x_{s}^2$. The length scale $x_s$ and the energy scale $E_s$ can be
chosen arbitrarily, but for a potential with power-law tail $V(x)\sim
C_nx^{-n}$, the van der Waals units with $x_s = (2\mu |C_n|/\hbar^2
)^{\frac{1}{n-2}}$ are most convenient. Thus, for the remainder of
this Letter, we redefine $x_s^2U(x) \to U(r)$ such that it is
dimensionless.  As an illustrative example, we use $U(r) =
-\varepsilon+V(r)$ with
\begin{equation}\label{eq:VCs2}
   V(r)= -\frac{1}{r^6} + V_0e^{-\gamma r},
\end{equation}
where $V_0=1.55\times 10^{12}$ and $\gamma=200$, which mimics the
a$^3\Sigma_u^{+}$ potential curve for Cs$_2$. The shaded area in
Fig.~\ref{fig:scheme}a between $r_1 = 0.052$ and $r_2=0.055$ marks the
optimization interval for $\varepsilon=-2\times 10^6$. The solutions
$\phi$ and $\chi$ are initialized at $r_1$ with $\phi=0,\ \phi'= k_1 =
\sqrt{-U(r_1)}$ and $\chi=1,\ \chi'=0$, such that $\phi$ and $\chi$
are similar to sine and cosine respectively. Thus a good choice for
the initial ansatz is $\rho_{\mathrm{ansatz}} = \phi^2+\chi^2$, shown
as the oscillatory red curve in Fig.~\ref{fig:scheme}b. Finally, we
minimize $\delta \rho$ and find the optimal values for $A,B,C$ which
give the smooth envelope $\rho_{\mathrm{opt}}(r)$ shown as the black
curve in Fig.~\ref{fig:scheme}b.  We emphasize that this procedure is
very robust with respect to the size of the optimization interval;
namely, we obtain the same values of $A,B,C$, when the interval is
enlarged to contain up to three oscillations.

Figure~\ref{fig:surf} compares our optimization procedure with the
standard WKB-initialized scheme \cite{Chris_Greene_QDT_surface}, which
relies on using the WKB approximation to impose the initial condition
for $\rho$ at the bottom of the potential ($r_0\approx 0.05856$). We
use Eq.~(\ref{eq:Milne-original-3rd-linear}) to compute both the
optimal envelope, $\rho_{\mathrm{opt}}(x)$, and WKB-initialized
envelope denoted as $\rho_{\mathrm{osc}}(x)$ for a range of energies
corresponding to $-7\times10^6<\varepsilon<4\times10^6$.  For clarity,
we make use of the WKB approximation $\rho_{\mathrm{wkb}}(r) \equiv
q|U(r)|^{-1/2}$ to rescale both $\rho_{\mathrm{opt}}(r)$ and
  $\rho_{\mathrm{osc}}(r)$. Thus, we define $Z_{\mathrm{opt}}$ and
  $Z_{\mathrm{osc}}$ according to
  $Z(\varepsilon,r)=\frac{\rho(r)}{\rho_{\mathrm{wkb}}(r)}-1$, and
  plot them in Fig.~\ref{fig:surf}. Note the oscillatory behavior of
  $Z_{\mathrm{osc}}$, while $Z_{\mathrm{opt}}$ obtained using our
  $\rho_{\mathrm{opt}}(r)$ is smooth.

The optimization method described above is applicable for all
classically allowed regions, and it provides a smooth envelope which
can be propagated efficiently by solving
Eq.~(\ref{eq:Milne-original-3rd-linear}). Note that when the
propagation enters a classically forbidden region, the envelope will
take on an increasing behavior, e.g., for $\varepsilon=-\kappa^2<0$
and $r\to \infty$, $y(r)\sim e^{\kappa r}$. Thus, the solution $\psi =
y(r)\sin \theta(r)$, with $\theta(r)=q\int_0^{r}
\frac{d\tau}{\rho(\tau)}$, will diverge when $r\to \infty$, unless
$\theta(\infty)$ is an integer number of $\pi$, in which case, $\sin
\theta(r) \sim e^{-2\kappa r}$ guarantees that $\psi$ has the correct
behavior of an eigenfunction corresponding to a bound state.  However,
for scattering solutions ($\varepsilon>0$), the asymptotic region
($r\to\infty$) can be tackled even more directly, as we show next.
\begin{figure}[b]
\includegraphics[width=1.0\columnwidth]{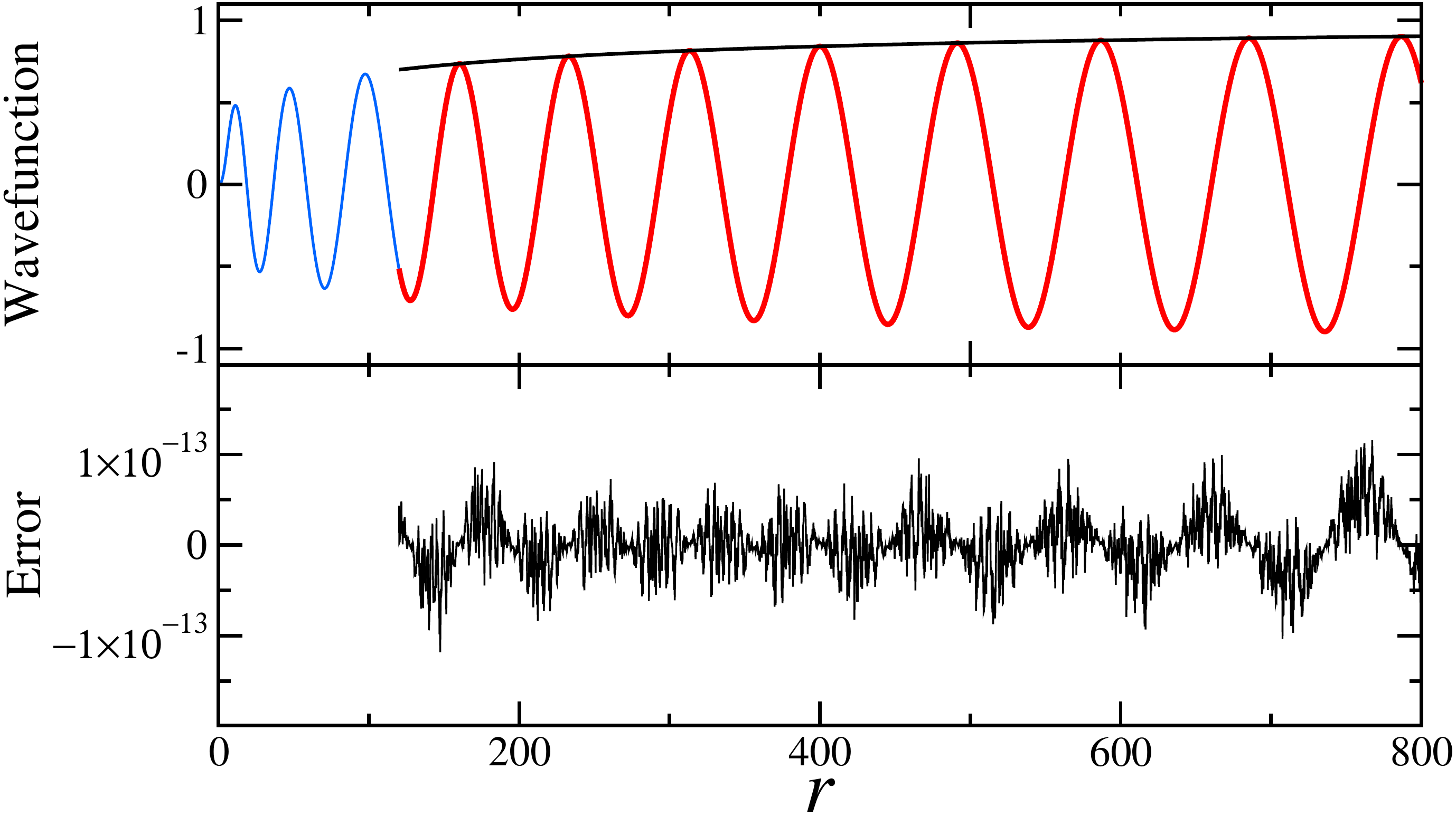}
\caption{\label{fig:wf_test} Accuracy test for an attractive Coulomb
  potential.  Upper: blue line for $F_\ell$ computed using the radial
  equation~(\ref{eq:DE-original}); red line for $F_\ell$ obtained from
  the envelope $\rho$ computed using Eqs.~(\ref{eq:rho-x-eq}) and
  (\ref{eq:Milne-original-3rd-linear}); black line for the amplitude
  $y=\sqrt\rho$.  Lower: the difference between the red and blue
  curves.  See text for details.}
\end{figure}

\textit{ Exact asymptotic solutions for scattering ---} For $E>0$, the
usual numerical approach for the radial Schr\"odinger equation
consists in propagating its solution far into the asymptotic region,
where matching is performed using Bessel or Coulomb functions, and the
S-matrix is extracted. Here we present a novel approach based on the
phase-envelope method for efficiently computing highly accurate
asymptotic solutions for any potential $V(r)$ which may contain
Coulomb and centrifugal terms.  First, the change of variable $z=1/r$
maps the infinite radial interval $[r_\mathrm{max},\infty]$ into a
finite interval $[0,z_\mathrm{max}]$ with
$z_\mathrm{max}=1/r_\mathrm{max}$, thus making it possible to enforce
boundary conditions at $r=\infty$ ($z=0$) and to account for the
entire tail of $V(r)$ without approximations. Secondly, the
\emph{linear} equation~(\ref{eq:Milne-original-3rd-linear}) allows for
a simple implementation of a spectral integration method employing a
small number of Chebyshev polynomials \cite{Greengard,Mihaila}, which
yields a highly accurate envelope.  Rewriting
Eq.~(\ref{eq:Milne-original-3rd-linear}) in the new variable $z$, we
have
\begin{equation}\label{eq:rho-x-eq}
z^4\partial_{z}^3 \rho+6z^3\partial_{z}^2\rho+6z^2\partial_{z}\rho - 4U\partial_{z}\rho -
2(\partial_{z}U)\rho=0.
\end{equation}
Next, in order to impose the initial condition $\rho|_{z=0}=1$,
Eq.~(\ref{eq:rho-x-eq}) is integrated only one time, and the smooth
envelope $\rho_{\infty}(z)$ is extracted as a unique solution of the
newly obtained equation, without the need for an explicit
optimization.  Indeed, all other possible solutions of
Eq.~(\ref{eq:rho-x-eq}) oscillate infinitely fast near $z=0$
($r=\infty$), and thus they exist outside the small subspace spanned
by the Chebyshev basis, which is restricted to polynomials of degree
$N\leq 20$.  The use of a small basis is of critical importance, as it
ensures a very effective suppression of oscillatory behavior, while it
is nevertheless sufficient for a highly accurate smooth solution.

As an illustrative example, we use an attractive Coulomb potential
with partial wave $\ell=2$ and $k=\sqrt{\varepsilon}=0.05$,
specifically, $U(r)=-1/r+6/r^2-k^2$.  For a Coulomb potential, it is
well known that the asymptotic phase also contains a logarithmic term,
which we separate explicitly:
$\theta_\infty(r)=kr+\frac{\ln(2kr)}{2k}-\frac\ell
2\pi+\tilde\theta(r)$.  Correspondingly, it is necessary to decompose
the envelope as $\rho_{\infty}(z)=1-\frac{z}{2k^2}+u(z)$. Hence the
nontrivial phase $\tilde{\theta}$ reads
  \begin{equation}\label{eq:Coulomb_asy_phase}
\tilde\theta(z) = k\int_0^z\frac{d\tau}{\tau^2}\frac{1}{\rho(\tau)}
 	\left[ u(\tau)+\frac{\tau}{2k^2}\left(u(\tau)-\frac{\tau}{2k^2}\right)\right], 
\end{equation}
which is computed using the Clenshaw--Curtis quadrature
\cite{Clenshaw_Curtis}. Note that we used $q=k$ in
Eq.~(\ref{eq:Coulomb_asy_phase}).

\begin{figure}[t]
\includegraphics[width=1.0\columnwidth]{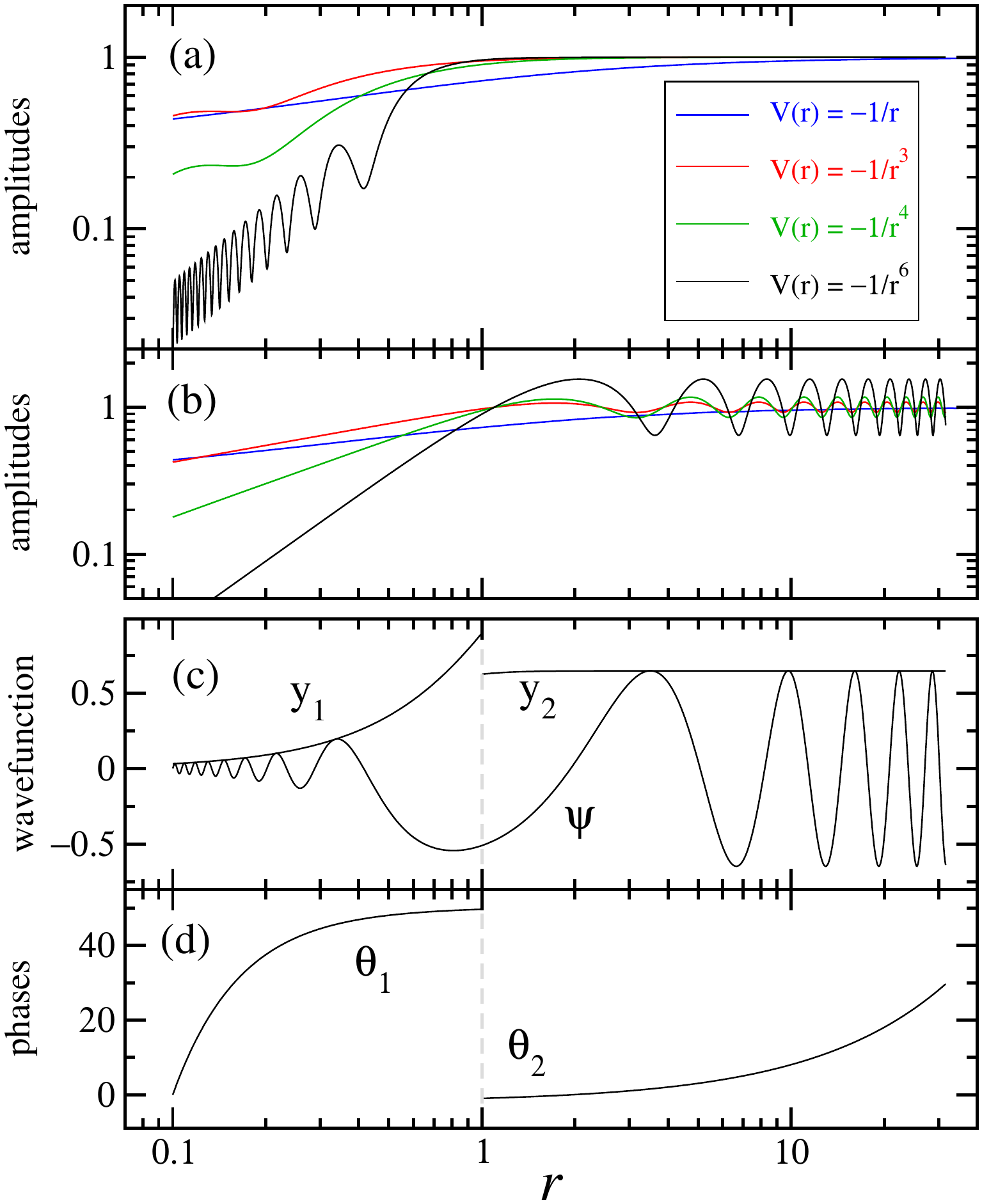}
\caption{(a) Asymptotically optimized amplitudes for $\varepsilon=1$ 
and $V(r)=-1/r^n$, with $n=1,3,4,6$. (b) Amplitudes
optimized at short range for the same cases.
(c) Regular solution $\psi=y_1\sin\theta_1=y_2\sin\theta_2$
for $V(r)=-1/r^6$.  (d) Phases $\theta_{1,2}$ corresponding
to amplitudes $y_{1,2}$.  See text for details.}\label{fig:quantum_reflection}
\end{figure}

  We emphasize that accurate solutions for the envelope and phase can
  be obtained in the interval $[0,z_\mathrm{max}]$ without the need to
  further subdivide it into smaller sectors.  Subsequently, it is
  convenient to revert to the original variable $r$, and then
  propagate the asymptotic envelope and phase \emph{inwards}, from
  $r_\mathrm{max}$ to $r=0$, according to
  Eqs.~(\ref{eq:Milne-original-3rd-linear}) and
  (\ref{eq:Milne-original-solution-y}).  The amplitude
  $y_{\infty}(r)=\sqrt{\rho_{\infty}(r)}$ and the phase
  $\theta_\infty(r)$ can now be used to construct any solution of the
  radial Schr\"odinger equation.  In particular, the regular solution
  obtained from the phase-envelope method can be compared with its
  version computed \emph{independently} as a solution of
  Eq.~(\ref{eq:DE-original}).  The former is expressed as
  $F_\ell(q,r)=y_{\infty}\sin(\theta_\infty+\sigma_\ell)$, where
  $\sigma_{\ell}$ is the Coulomb phase shift, while the latter is
  initialized at $r=0$ with the appropriate behavior, $F_\ell\sim
  r^{\ell+1}$, and a suitable normalization factor
  \cite{Seaton_and_peach,George}, and then it is propagated
  \emph{outwards}.  As shown in Fig.~\ref{fig:wf_test}, the two
  independently computed wavefunctions agree to thirteen digits,
  nearly the full fifteen digits available in standard
  double-precision computer arithmetic.  We emphasize that, apart from
  the appropriate factors used to initialize the solutions of
  Eqs.~(\ref{eq:DE-original}) and (\ref{eq:rho-x-eq}), no
  rescaling of the wavefunctions was necessary for the comparison in
  Fig.~\ref{fig:wf_test}.

\textit{ Combining locally optimized solutions ---} When two
classically allowed regions are separated by a classically forbidden
region due to a barrier, it is well known \cite{Multiple_well_Soo_Yin}
that a global envelope which is smooth in all regions cannot exist. In
fact, this lack of global smoothness can also manifest within a single
classically allowed region.  Indeed, when the asymptotically optimized
envelope is propagated inwards, it may develop oscillations at short
range, as shown in Fig.~\ref{fig:quantum_reflection}a. Conversely, if
the envelope is first optimized at short range, it may develop
oscillations when propagated outwards into the asymptotic region, see
Fig.~\ref{fig:quantum_reflection}b. This type of oscillatory behavior
is directly related to quantum
reflection\cite{Harald_2004,Robin_Harald,Robin_qr_mirror,Robin_qr_mirror2,
  Robin_qr_mirror3,Robin_qr_mirror4}, which is very pronounced at low
energy, but diminishes and eventually disappears at high energy.  The
results shown in Figs.~\ref{fig:quantum_reflection}(a) and (b)
correspond to $\ell=0$ and $V(r)=-\frac{1}{r^n}$ with
$n=1,3,4,6$. Note that the oscillations are more pronounced for high
$n$, due to the more abrupt behavior of $V(r)$, while $n=1$ (Coulomb)
is a special case which admits a globally smooth envelope for all
energies.

In the absence of a globally smooth envelope, a simple partitioning
scheme can be used to take advantage of regionally smooth phases and
envelopes. To illustrate such an approach, we use $V(r)=-1/r^6$ with
$\ell=0$ and $\varepsilon=1$. In the first region ($r \leq 1$) we
employ the short range optimization to obtain $\psi = y_1\sin
\theta_1$, with $\theta_1 = 0$ at $r_0=0.1$, where we placed a hard
wall. In the asymptotic region ($r \geq 1$) we construct the solution
$\psi=y_2\sin \theta_2$ with $y_2 = cy_{\mathrm{\infty}}$ and
$\theta_2 = \theta_{\infty}+\delta$. Matching for $\psi$ and $\psi'$
is imposed at $r=1$ to determine $c$ and $\delta$.  The amplitudes
$y_1(r)$ and $y_2(r)$ are shown in Fig.~\ref{fig:quantum_reflection}c
and the phases $\theta_1(r)$ and $\theta_2(r)$ are shown in
Fig.~\ref{fig:quantum_reflection}d. Note that $\theta'_1\neq\theta'_2$
and $\theta_1\neq\theta_2$ (mod $\pi$) at $r=1$.  Thus, despite
quantum reflection, any wavefunction $\psi(r)$ can still be
parametrized economically by judiciously partitioning the $r$ domain
and computing separately a smooth envelope and the corresponding phase
for each region.


\textit{ Conclusions ---} In this Letter, we derived a linear
alternative, Eq.~(\ref{eq:Milne-original-3rd-linear}), to Milne's nonlinear equation, and presented new
approaches for solving Milne's amplitude equation. For short and
medium range, we developed a simple and practical recipe for
optimization which provides a smooth envelope. For $E>0$ the
non-oscillatory envelope and phase can be computed very easily for the
entire asymptotic region for any type of potential.  In turn, the
optimized envelope allows for the computation of quantities which have
a smooth energy dependence.  We also showed that very high numerical
precision can be obtained in a straightforward manner using the
prescribed approaches. Finally, extension of this method to coupled
channel problems is underway, where the applicability of
Chebyshev-based solvers using the mapping $z=\frac{1}{r}$,
is explored.

This work was partially supported by the MURI US Army Research Office
Grant No. W911NF-14-1-0378 (IS) and by the US Army Research Office,
Chemistry Division, Grant No. W911NF-13-1-0213 (DS and RC).

\bibliography{Milne}

\end{document}